\documentclass{PoS}

\usepackage{graphicx}
\usepackage{amsmath,amssymb,amsfonts}
\usepackage{subfigure}
\usepackage{listings}
\usepackage{cite}

\newcommand{\GOSAM}{{\textsc{Go\-Sam}}}

\newcommand{\QGRAF}{{\texttt{QGRAF}}}
\newcommand{\FORM}{{\texttt{FORM}}}
\newcommand{\SPINNEY}{{\texttt{spin\-ney}}}
\newcommand{\HAGGIES}{{\texttt{hag\-gies}}}
\newcommand{\SAMURAI}{{\textsc{Sa\-mu\-rai}}}

\newcommand{\bea}{\begin{eqnarray}}
\newcommand{\eea}{\end{eqnarray}\noindent}

\newcommand{\bcen}{\begin{center}}
\newcommand{\ecen}{\end{center}}
\def\url#1{\texttt{#1}}

\def\Gosam{{{\sc GoSam}}}
\def\gosam{{{\sc GoSam}}}

\def\samurai{{{\sc samurai}}}
\def\Sherpa{{{\sc Sherpa}}}

\def\C++{{{\sc c++}}}

\def\QCDLoop{{{\sc QCDLoop}}}
\def\OneLoop{{{\sc OneLoop}}}
\def\Golem{{{\sc Golem95C}}}

\def\Ninja{{{\sc Ninja}}}

\def\QCDLoop{{{\sc QCDLoop}}}
\def\tth{{$t\bar{t}H$}}
\def\tthj{{$t\bar{t}Hj$}}

\title{GoSam @ LHC: algorithms and applications to Higgs production }

\ShortTitle{GoSam @ LHC}

\author{G.~Cullen\\
Deutsches Elektronen-Synchrotron DESY, Platanenallee 6, 15738 Zeuthen, Germany\\
E-mail: \email{gavin.cullen@desy.de}}

\author{H.~van Deurzen, N.~Greiner, G.~Heinrich,  G.~Luisoni, E.~Mirabella,
T.~Peraro, J.~Reichel, J.~Schlenk, J.F. von Soden-Fraunhofen\\
Max Planck Institute for Physics, F\"ohringer Ring 6, 80805 Munich, Germany\\
E-mail: \email{\{hdeurzen, greiner, gudrun, luisonig, mirabell, peraro, joscha, jschlenk, jfsoden\}@mpp.mpg.de}}

\author{\speaker{P.~Mastrolia}\\
Max Planck Institute for Physics, F\"ohringer Ring 6, 80805 Munich, Germany; \\ Dipartimento di Fisica e Astronomia, Universit\`a di Padova, and INFN Sezione di Padova,
via Marzolo 8, 35131 Padova, Italy\\
 E-mail: \email{ppaolo@mpp.mpg.de}}

\author{G.~Ossola\\
Physics Department, New York City College of Technology, The
City University of New York,
300 Jay Street Brooklyn, NY 11201, USA; \\
The Graduate School and University Center, The City University of New York,
365 Fifth Avenue, New York, NY 10016, USA\\
 E-mail: \email{gossola@citytech.cuny.edu}}

\author{F.~Tramontano\\
Dipartimento di Scienze Fisiche, Universit\`a degli studi di Napoli and INFN, Sezione di Napoli,
80125 Napoli, Italy\\
E-mail: \email{francesco.tramontano@cern.ch}}

\abstract{
We elaborate on {\gosam}, a code-writer for
automated one-loop calculations. After recalling its main features, we
present a selection of phenomenological results
recently obtained, giving relevance at the evaluation of NLO QCD
corrections to the production of a Higgs boson
in association with jets and heavy quarks.
}

\FullConference{11th International Symposium on Radiative Corrections (Applications of Quantum Field Theory to Phenomenology) (RADCOR 2013),\\
		22-27 September 2013\\
		Lumley Castle Hotel, Durham, UK }

\begin{document}

\section{Introduction}

Automating the evaluation of Next-to-leading order (NLO) QCD
corrections for high multiplicity final-state processes was considered beyond the horizon of possibilities only a few
years ago.
The overwhelming complexity of one-loop Feynman integrals seemed to
prevent from devising a single algorithm which could systematically
be applied to {\it all} processes with a same theoretical and computing effort.
Indeed, a process-by-process strategy seemed the only viable
option to pursue, and final answers could be obtained by collecting the
results of several calculations, each tackling individual contributions of
increasing complexity.
The analytic calculation of the one-loop six-gluon amplitudes
\cite{Bern:1994zx,Bern:1994cg,Britto:2005ha,Bidder:2004tx,Bidder:2004vx,Bidder:2005ri,Bern:2005hh,Bedford:2004nh,Britto:2006sj,Xiao:2006vt}
and the one-loop Higgs plus four parton amplitudes
\cite{Berger:2006sh,Badger:2006us,Badger:2007si,Glover:2008ffa,Badger:2009hw,Badger:2009vh,Dixon:2009uk}
are representative examples of these long-lasting theoretical {\it enterprises}.
But already the completion of these two calculations contained
the seeds of a change of perspective, and
in less than a decade, the developments of new analytic
insights on the structure of scattering amplitudes made the automation possible.

The role played by the investigation of the integrand structures of
Feynman amplitudes~\cite{Ossola:2006us}, in particular the study of
their singularities \cite{Britto:2005fq} and
the deep understanding of the corresponding residues \cite{Britto:2004nc}
had a dramatic impact on the development of efficient algorithms for the
quantitative determination of one-loop amplitudes.
This goal was achieved by combining unitarity and analyticity of
scattering amplitudes~\cite{Bern:1994cg,Bern:1994zx}  together with the biunivocal relation
established between the integrand decomposition \cite{Ossola:2006us} and the expression of
the integrated results in terms of a independent set of basic
integrals, addressed to as {\it master integrals} (MI's).
The availability of analytic results for MI's for {\it any} (in
principle) partonic one-loop process, let the community focus on the
development of efficient algebraic methods for the determination of
the coefficients multiplying them.

The breakthrough was achieved by conjugating three pieces of information:
one-loop amplitudes factorize in the product of simpler, tree-level amplitudes when
multiparticle singularities are approached;
complex momenta can be exploited for imposing on-shell cut-conditions on
propagating particles; residues at the singular configurations are polynomials in those components of the loop
momenta which are not constrained by the on-shell conditions. The form
of such polynomials is process independent, while the values of their
coefficients depend on the particle content and their interactions.
The insurmountable problem of integrating one-loop diagrams turned
into the algebraic problem of determining polynomial coefficients.

The overcoming of the bottleneck represented by the evaluation of
one-loop virtual corrections found the proper tandem in the optimizations
of algorithms dedicated to the event generation at NLO accuracy, which
usually provide the complementary contributions of phase-space integration
together with a subtraction procedure for the treatment of infrared singularities.

Meanwhile, the data collected by the experimental
collaborations at the Large Hadron Collider (LHC) has been allowing for a
detailed investigation of the Standard Model (SM) of particle physics,
culminating in the exciting confirmation of the validity of the
electroweak symmetry breaking
mechanism~\cite{Englert:1964et,Higgs:1964ia} with
the discovery of a scalar boson with mass of about 126 GeV~\cite{Aad:2012tfa,Chatrchyan:2012ufa}.

To further study the properties of the recently discovered Higgs boson, theory predictions play a fundamental role. They are not only needed for the signal, but also for the modeling of the relevant background processes, which share similar experimental signatures. Further, precise theory predictions are important in order to constrain model parameters in the event that a signal of New Physics is detected.
Since leading-order (LO) results are affected by large uncertainties, theory predictions are not reliable without accounting for higher orders. Therefore, it is of primary interest to provide theoretical tools which are able to perform the comparison of LHC data to theory  at NLO accuracy.

In this communication, we recall the main features of {\gosam}
~\cite{Cullen:2011ac}, a code writer for the automated computation of
one-loop amplitudes. {\gosam} has been recently employed in several
calculations at NLO QCD accuracy related to signal and backgrounds for
Higgs boson production~\cite{Greiner:2011mp,Greiner:2012im,
  vanDeurzen:2013rv,
  Gehrmann:2013aga,Gehrmann:2013bga,Cullen:2013saa,vanDeurzen:2013xla,
  Dolan:2013rja}, as well as in the context of Beyond Standard Model
(BSM) scenarios~\cite{Cullen:2012eh,Greiner:2013gca} and electroweak
studies~\cite{Chiesa:2013yma, Mishra:2013una},
and has been successfully interfaced with Monte Carlo programs to
merge multiple NLO matrix elements with parton showers~\cite{Luisoni:2013cuh, Hoeche:2013mua}.

We also briefly describe a selection of recent phenomenological
results obtained with {\gosam}, with particular attention to the
recent calculations of NLO QCD corrections to the production of a Higgs
boson in association with jets and heavy-quarks at the LHC, which
stimulated some of the technical advances that will go in the
forthcoming release of our code.

\section{ \GOSAM{}: the framework}

{\gosam} combines automated diagram generation and algebraic manipulation~\cite{Nogueira:1991ex, Vermaseren:2000nd, Reiter:2009ts, Cullen:2010jv} with integrand-reduction
techniques~\cite{Ossola:2006us, Ossola:2007bb, Ellis:2007br, Ossola:2008xq,  Mastrolia:2008jb,Mastrolia:2012bu}.
Amplitudes are generated via Feynman diagrams, using \QGRAF~\cite{Nogueira:1991ex}, \FORM~\cite{Vermaseren:2000nd}, \SPINNEY~\cite{Cullen:2010jv} and \HAGGIES~\cite{Reiter:2009ts}.
The role of the individual programs are managed by python scripts, so that
the only task required from the user is the preparation of an input
file, needed for launching the generation of the source code and its
compilation.
The input file contains specific information about:
{\it i.} the {\it
  process} (initial and
      final state particles, the order in the coupling constants, and the model);
{\it ii.} the {\it scheme} (regularization and renormalization schemes);
{\it iii.} the {\it system} (libraries or compiler options);
{\it iv.} additional options to control/optimize the code generation.

After the generation of all contributing diagrams, the virtual corrections are evaluated using the $d$-dimensional integrand-level reduction method,
as implemented in the library \SAMURAI~\cite{Mastrolia:2010nb}, which allows for the combined
determination of both cut-constructible and rational terms at once.
Alternatively, the tensorial decomposition provided by
{\Golem}~\cite{Binoth:2008uq,Heinrich:2010ax,Cullen:2011kv}  is also
available. Such reduction, which is numerically stable but more time
consuming, is employed as a rescue system.
After the reduction, all relevant master integrals can be computed by
means of
{\Golem}~\cite{Cullen:2011kv},
{\QCDLoop}~\cite{vanOldenborgh:1990yc, Ellis:2007qk}, or {\OneLoop}~\cite{vanHameren:2010cp}.

\GOSAM{} can be used to generate and evaluate one-loop corrections in both QCD and electro-weak theory.
Model files for BSM theories can be generated from a Universal FeynRules Output (\texttt{UFO})~\cite{Christensen:2008py,Degrande:2011ua, Alloul:2013bka} or
\texttt{LanHEP}~\cite{Semenov:2010qt} file.

\subsection{Code development} \label{code}

\paragraph{Diagram Generation.}
New features have been recently implemented within \GOSAM{}, with respect  to the current public version.
In order to deal with the complexity level of calculations such as $pp\to Hjjj$~\cite{Cullen:2013saa}, the {\gosam} code has been enhanced. On the one side, the generation algorithm has been improved by a more efficient diagrammatic layout: Feynman diagrams are grouped according to their
topologies, namely global numerators are constructed by combining
diagrams that have a common set, or subset, of denominators,
irrespectively of the specific particle content.  On the other side,
additional improvements in the performances of {\sc GoSam} have been
achieved by exploiting the optimized manipulation of polynomial
expressions available in \FORM~4.0~\cite{Kuipers:2012rf}.
The possibility of employing numerical polarization vectors and the
option to sum diagrams sharing the same propagators algebraically
during the generation of the code led to an enormous gain in code
generation time and in reduction of code size.

\paragraph{Integrand Reduction.}
In~\cite{Mastrolia:2012bu}, some of us proposed a novel approach to the
integrand reduction, based on the determination master integral
coefficients by means of Laurent expansion performed by {\it polynomial divisions}.
This method has been implemented in the \C++ library {\Ninja},
and interfaced to the {\gosam} framework \cite{NinjaGo}
showing an improvement in the computational performance, both in terms
of speed and precision, with respect to the standard
algorithms. Further details can be found in the contribution of T.~Peraro at this conference~\cite{TizianoRC13}. The new library has been recently employed in the evaluation of NLO QCD corrections to $p p \to t {\bar t} H j $~\cite{vanDeurzen:2013xla}.

Concerning the amplitude reduction, {\sc GoSam} has been enhanced to
reduce integrands that may exhibit numerators with rank larger than
the number of the denominators. This is indeed the case, for instance,
when computing the Higgs boson production in gluon fusion within the
large top-mass approximation~\cite{vanDeurzen:2013rv,
  Cullen:2013saa}, or when dealing with spin-2 particles~\cite{Greiner:2013gca}.
For these cases, within the context of integrand-reduction techniques, the parametrization of the residues at the
multiple-cut has to be extended and the decomposition of any one-loop
amplitude acquires new master integrals~\cite{Mastrolia:2012bu}. The
extended integrand decomposition has been implemented in the \samurai\ library~\cite{Mastrolia:2012du}.

\paragraph{MC Interfaces and BLHA.}
The computation of physical observables at NLO accuracy, such as cross sections and differential distributions, requires to combine the one-loop results for the virtual amplitudes obtained with \gosam{}, with a Monte Carlo (MC) framework, that can take care of the phase-space integration and of the combination of the different pieces of the calculation. The communication between the programs is achieved by a standard interface, dubbed \emph{Binoth Les Houches Accord} (BLHA)~\cite{Binoth:2010xt,Alioli:2013nda}. More details are provided in the contribution of G.~Luisoni at this conference~\cite{GionataRC13}. \\

The new developments regarding the improved generation and reduction algorithms, as well as including the update BLHA standard, will be publicly available in the forthcoming release of \GOSAM\,2.0.

\section{Higgs boson production in Gluon Fusion}

Higgs production via gluon-fusion (GF) is one of the phenomenologically most relevant Higgs boson production process at the LHC. Indeed, not only it is the dominant production channel, but it constitutes an irreducible background for the production via vector-boson-fusion (VBF). The latter allows one to extract information about the couplings among the Higgs and the gauge bosons, and therefore to directly probe the symmetry breaking mechanism. In phenomenological analysis the pollution of GF events is reduced by applying a veto on the jet activity in the central region of the detector~\cite{DelDuca:2001eu,DelDuca:2001fn}. Therefore precise predictions for Higgs plus jets production in GF is essential for an accurate estimation of the jet-veto efficiency.

The computation of Higgs production via GF is theoretically challenging since the LO contribution is mediated by a heavy quark loop~\cite{DelDuca:2001eu,DelDuca:2001fn}. In these articles it was shown that the large top-mass limit~\cite{Dawson:1990zj} is a reliable approximation when the jet transverse momenta are smaller than the top quark mass $m_t$, allowing one to compute predictions at NLO accuracy in this limit~\cite{Dittmaier:2011ti, Dittmaier:2012vm, Heinemeyer:2013tqa}.
% At the LHC, the dominant Higgs production mechanism proceeds via gluon fusion (GF), where the coupling of the Higgs to the gluons is mediated by a heavy quark loop.

%For this reason, the calculation of higher order corrections for the GF production of a Higgs boson in association with jets has received a lot of attention in the theory community over the past decade~\cite{Dittmaier:2011ti, Dittmaier:2012vm, Heinemeyer:2013tqa}.

\begin{figure}[h]
\hspace{1.0cm}
\subfigure[]{\includegraphics[width=6cm]{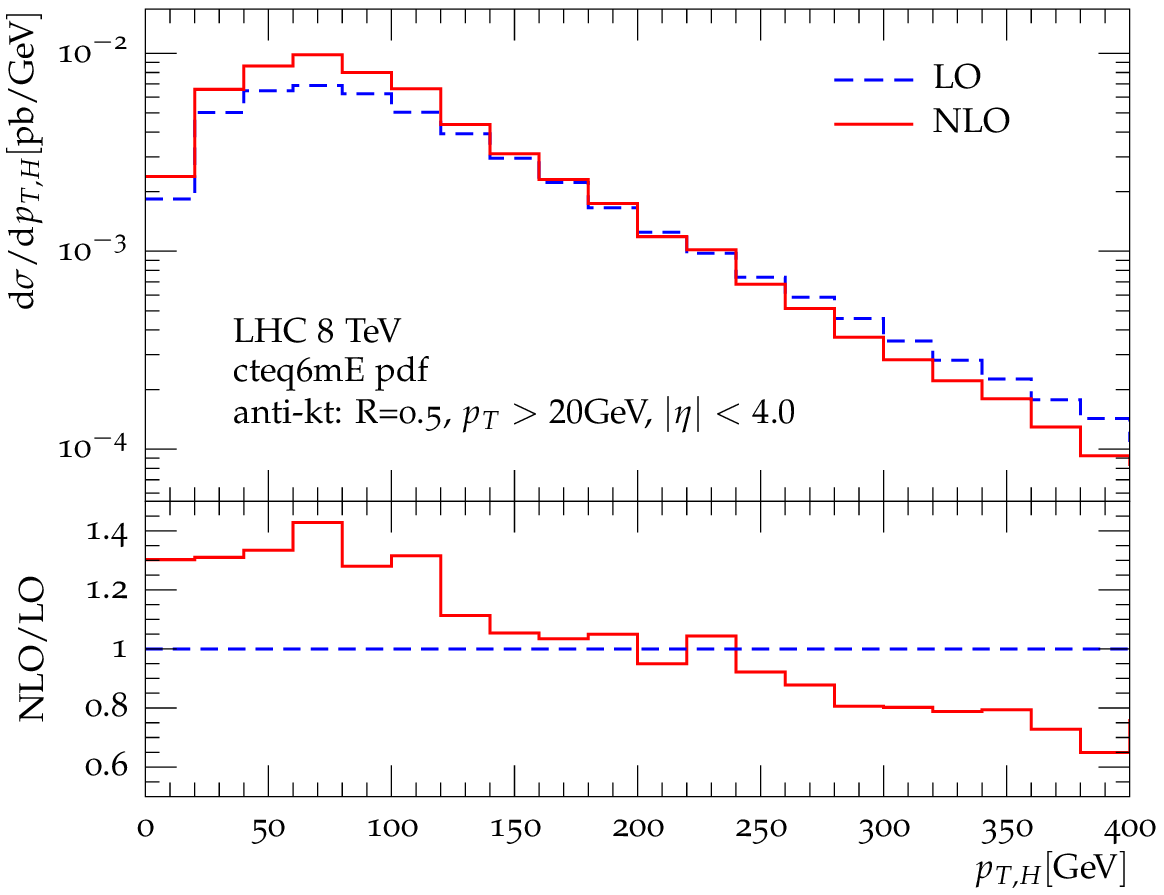}}
\hspace{1.0cm}
\subfigure[]{\includegraphics[width=6.8cm]{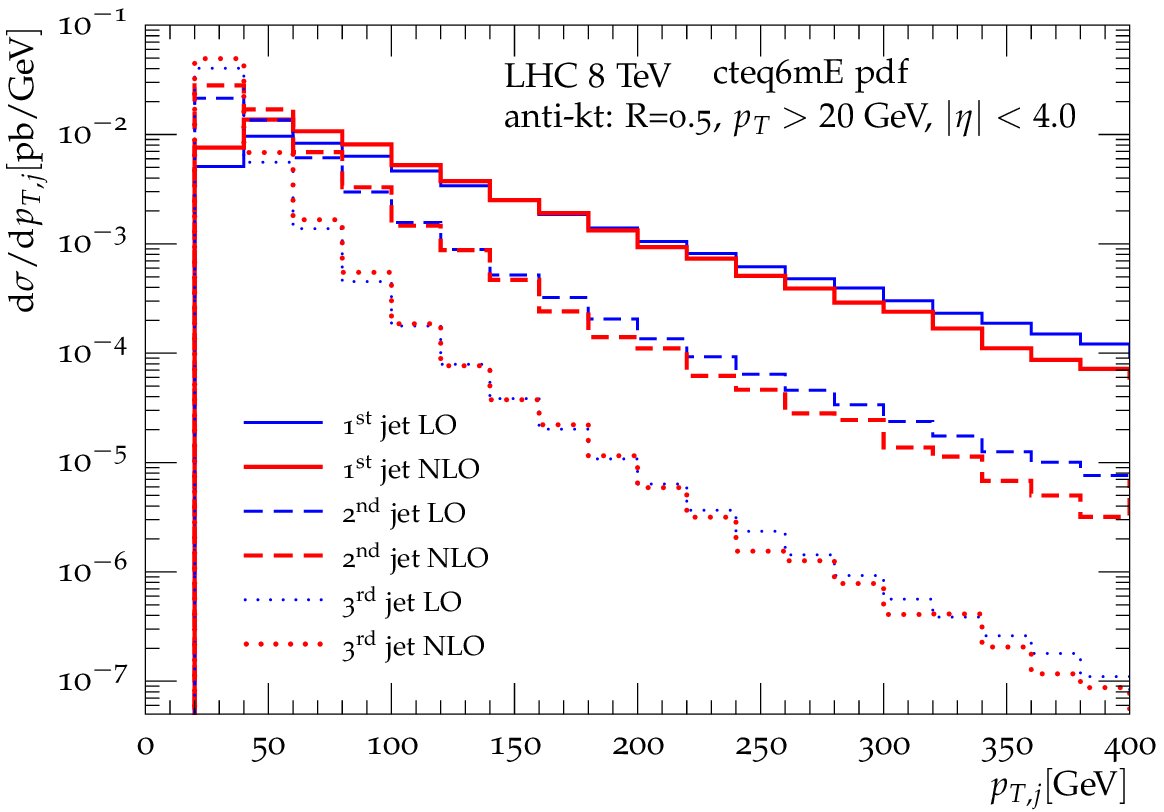}}
\caption{ $pp \to H+3 j$ in GF for the LHC at 8 TeV: (a) transverse momentum distributions of the leading jets, (b) transverse momentum distribution of the
  Higgs boson.}
\label{fig:H3j}
\end{figure}

The developments in \GOSAM{}, described in Section~\ref{code}, allowed us to compute the NLO QCD corrections to the production of $H+2$ jets $(Hjj)$~\cite{vanDeurzen:2013rv} and, for the first time, also $H+3$ jets $(Hjjj)$~\cite{Cullen:2013saa} in GF in the large top-mass limit.

While a fully automated BLHA interface between \GOSAM{} and \Sherpa{}~\cite{Gleisberg:2008ta} has been used for $Hjj$, the complexity of the integration for the process $Hjjj$ forced us to employ a hybrid setup which combines \GOSAM{}, \Sherpa{} and the MadDipole/Madgraph4/MadEvent framework~\cite{Frederix:2008hu,Frederix:2010cj,Stelzer:1994ta,Maltoni:2002qb,Alwall:2007st}.
This calculation is indeed challenging both on the side of real-emission contributions and of the virtual corrections, which alone involve more than $10,000$ one-loop Feynman diagrams with up to rank-seven hexagons.

In both calculations the cteq6L1 and cteq6mE parton-distribution functions were used for LO and NLO respectively, and a minimal set of cuts based on the anti-$k_T$ jet algorithm with
$R=0.5$, $p_{T,min}>20$ GeV and $\left|\eta\right|<4.0$ was applied.

In the case of $Hjjj$ the transverse momentum distribution of the Higgs boson and the three leading jets are shown in Figure~\ref{fig:H3j}. In all the distributions the NLO corrections enhance the LO prediction in the low $p_T$ region ($p_T \lesssim 150,200$ GeV), whereas their contribution is negative at higher $p_T$.

This study also shows that the virtual contributions for $Hjjj$ generated by {\Gosam} is ready to be paired with available Monte Carlo programs to aim at further phenomenological studies.

\section{Associated Higgs boson production with a top-pair and a jet}

Together with the GF channel, the production of a Higgs boson in association with a pair of top quarks (\tth) permits to access the Yukawa coupling between the Higgs boson and the top quark. In particular, selected distributions could shed light on the coupling structure and on $CP$ properties of the Higgs boson. Experimentally this channel is difficult to measure because of the large $t\bar{t}$ plus (light or heavy flavour) jets background and the combinatorial background. Recent studies have shown that sensibility to \tth can be improved either by considering boosted top quarks and Higgs boson in the final state~\cite{Plehn:2009rk} or by using the matrix element method~\cite{Artoisenet:2013vfa}.

%The production rate for a Higgs boson associated with a top-antitop pair ($t \bar t H$) is particularly interesting to study the properties of the newly discovered Higgs boson, since it is directly proportional to the SM Yukawa coupling of the Higgs boson to the top quark.
Recently we presented the complete NLO QCD corrections to the process $ pp \to t \bar t H + 1$ jet  (\tthj) at the LHC~\cite{vanDeurzen:2013xla}. The goal of the calculation was twofold.  Besides its importance for the phenomenological analyses at the LHC, this process is also interesting from a technical point of view. Indeed the presence of two mass scales (Higgs boson and top quark) and of internal massive particles affects the stability of many numerical reduction algorithms. This calculation represents the first application of the novel reduction algorithm, implemented in the library {\Ninja},  based on integrand-level reduction via Laurent expansion~\cite{Mastrolia:2012bu}.
%The new algorithm performed very well both in terms of stability and speed.

\begin{figure}[h]
\hspace{1.0cm}
\subfigure[]{\includegraphics[width=7.0cm]{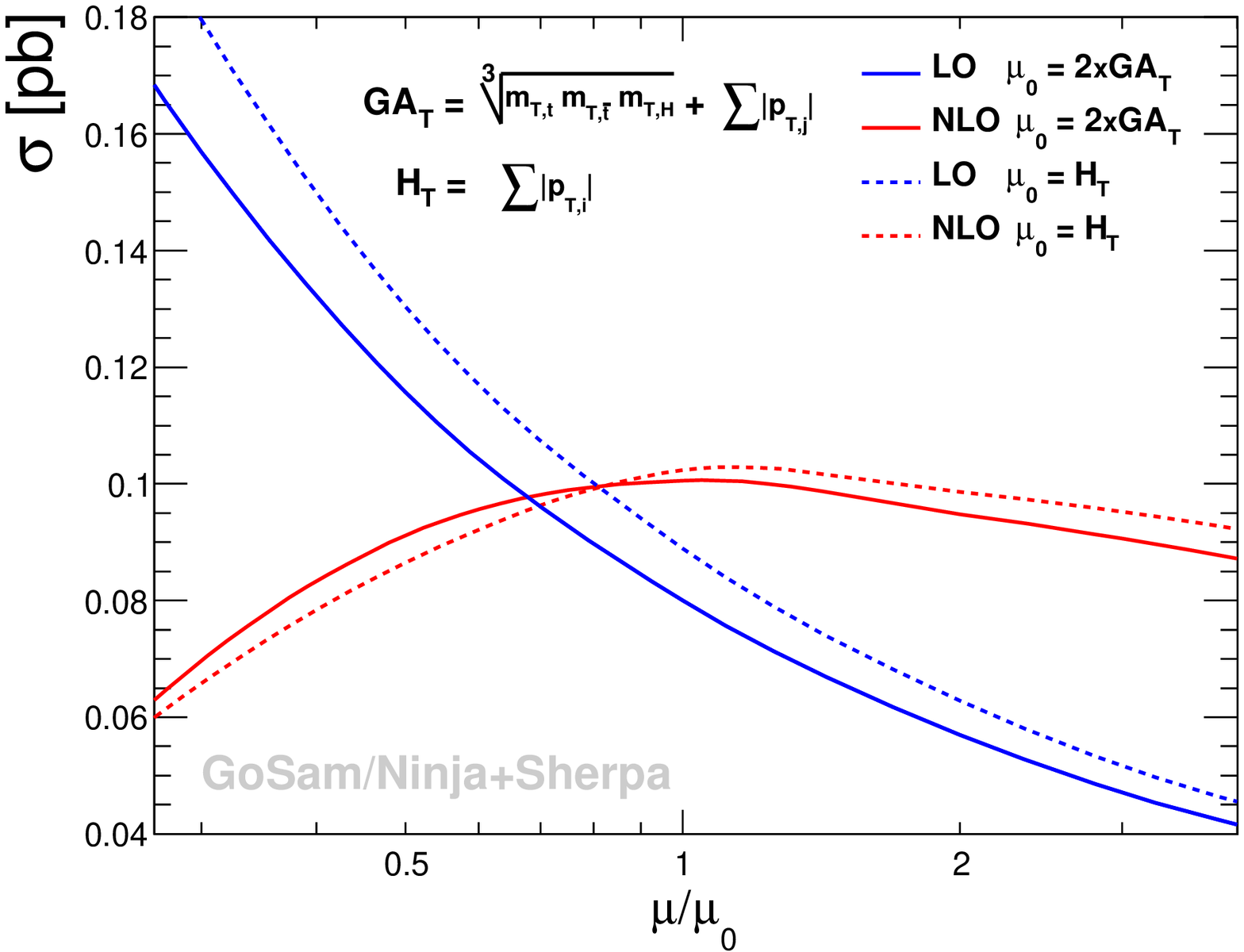}}
\hspace{1.0cm}
\subfigure[]{\includegraphics[width=6cm]{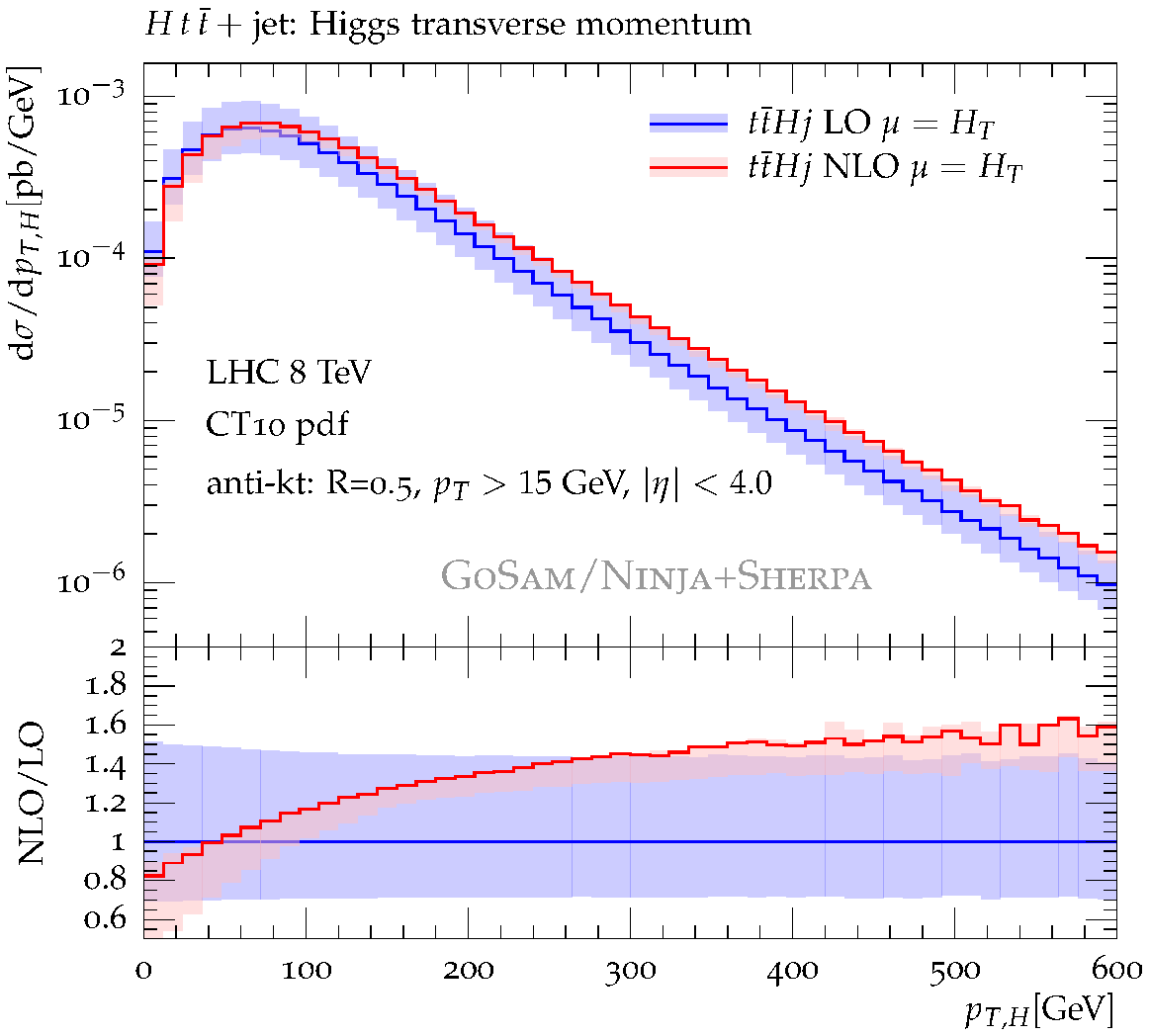}}
\caption{ $pp \to t\bar{t }H + j$ for the LHC at 8 TeV: (a) scale dependence of the total cross section for two values of the central scale, (b) transverse momentum distributions of the Higgs boson.}
\label{fig:tthj}
\end{figure}

In Figure~\ref{fig:tthj} we present the impact of NLO corrections to \tthj for the LHC at 8 TeV. The jets are clustered using the anti-$k_t$-algorithm with radius $R=0.5$, a minimum transverse momentum of $p_{T,jet}>15$ GeV and pseudorapidity $|\eta|<4.0$. The LO cross sections are computed with the LO parton-distribution functions cteq6L1~\cite{Pumplin:2002vw},
whereas at NLO we use CT10~\cite{Lai:2010vv}.

The scale dependence of the total cross section (Fig.~\ref{fig:tthj}a) is strongly reduced by the inclusion of the NLO contributions. It is worthwhile to notice that both choices for the central value of the scale, defined in the plot, provide an adequate description, being close to the physical scale of the process. In particular the scale choice $\mu\simeq0.8\,H_T$ minimizes the impact of the NLO corrections, improving the reliability of the LO prediction for the production rate. The transverse momentum distribution of the Higgs boson, $p_{T,H}$, is shown in Fig.~\ref{fig:tthj}b. The numerical impact of the NLO corrections increases at $p_{T,H}$ increases, and is of the order of $50-60\%$ of the LO prediction for boosted Higgs bosons ($p_{T,H}\gtrsim 400$ GeV).

\section{Other Phenomenological results}

%\paragraph{Diphotons and jets.}

\GOSAM{} in combination with MadDipole/MadGraph4/MadEvent has been
used to calculate the NLO QCD corrections to $pp\to\gamma\gamma +1,2$
jets~\cite{Gehrmann:2013aga ,Gehrmann:2013bga} and to the production
of a graviton in association with one jet~\cite{Greiner:2013gca},
where the graviton decays into a photon pair, within ADD models of
large extra dimensions~\cite{ArkaniHamed:1998rs,
  Antoniadis:1998ig}. Furthermore it has been used to compute the NLO
Susy-QCD corrections to $pp\to \chi_{1}^{0}\chi_{1}^{0}+1$
jet~\cite{Cullen:2012eh}. \\
{\Gosam} was also interfaced with the POWHEG BOX to compute $pp\to
HV+1$~jet ($V=W^{\pm},\,Z$) \cite{Luisoni:2013cuh}, and with Sherpa to
calculate the NLO QCD corrections to 
$pp\to W^{+}W^{-}b\bar{b}$, where the $W$ bosons decay leptonically~\cite{JohannessRC13}. \\
Finally \Gosam{} was also used to compute the production of a pair of
 Higgs bosons in association with two jets in~\cite{Dolan:2013rja}.

See the contributions of N.~Greiner, G.~Luisoni and J.~Schlenk~\cite{NicoRC13,GionataRC13,JohannessRC13}
for dedicated discussions on these topics. 

\section{Conclusions}

\GOSAM{} is a flexible and widely applicable tool for the automated calculation of multi-particle scattering one-loop amplitudes. After interfacing it with MC programs, that can perform integration over phase space and combine the contributions coming from real emission and subtraction terms as well, total cross-sections and differential distributions at NLO accuracy can be easily obtained for a variety of processes of interest at the LHC. %Recent examples of calculations performed with \GOSAM{} include NLO QCD corrections to Higgs production channels and backgrounds, neutralino and graviton production in BSM scenarios, and also some electroweak corrections.

Boosted by state-of-the-art techniques for the reduction of the scattering amplitudes,  \GOSAM{} provides a reliable answer for multi-leg amplitudes in the presence of massive internal and external legs and propagators, such as the production of a Higgs boson in conjunction with a top-quark pair, as well as in configurations with relatively high multiplicity, such as Higgs boson plus jets production.
While the \GOSAM{} code will be further improved, it will be
interesting to observe whether the extension of integrand-level
techniques~\cite{Mastrolia:2011pr, Mastrolia:2012an,Mastrolia:2012wf, Mastrolia:2013kca} to higher orders will succeed and provide a
comparable level of automation, at least for the calculation of the
virtual parts.

Other challenges for the near future involve interfacing \GOSAM{} with MC programs for an automated generation of the full cross section including parton showering, and ultimately the production of codes and results to be used within experimental analyses. We believe that the amount of recent calculations that were produced with the \GOSAM{} framework shows, both in terms of stability and precision, that it is an ideal multi-purpose tool for studying the physics at the LHC. %Further optimizations and interfaces are currently in progress and will be part of the future \GOSAM\,2.0  public release of the code.

\subsection*{Acknowledgments}
The work of G.C. was supported by DFG SFB Transregio 9, and by Research Executive Agency (REA) of the European Union
under the Grant Agreement number PITN-GA-2010-264564 (LHCPhenoNet). H.v.D., G.L., P.M., and T.P. are supported by the Alexander
von Humboldt Foundation, in the framework of the Sofja Kovaleskaja Award Project, %“Advanced Mathematical Methods for Particle Physics”,
endowed by the German Federal Ministry of Education and Research. The work of G.O. was supported in part by the National Science Foundation
under Grant PHY-1068550 and PSC-CUNY Award No. 65188-00 43.
This research work benefited of computing resources from the Rechenzentrum Garching and the CTP cluster of the New York City College of Technology.
%\clearpage

%\section*{References}
\bibliographystyle{iopart-num}
\bibliography{references}

%\providecommand{\newblock}{}
%\begin{thebibliography}{10}
%\expandafter\ifx\csname url\endcsname\relax
%  \def\url#1{{\tt #1}}\fi
%\expandafter\ifx\csname urlprefix\endcsname\relax\def\urlprefix{URL }\fi
%\providecommand{\eprint}[2][]{\url{#2}}

\end{document}